\documentclass[prl,floatfix,twocolumn,showpacs]{revtex4}

\usepackage{graphicx}%
\usepackage{dcolumn}
\usepackage{amsmath}
\usepackage{amssymb}
\begin{document}


\title{Robust Emergent Activity in Dynamical Networks}
\author{Sitabhra Sinha and Sudeshna Sinha}
\affiliation{
 The Institute of Mathematical Sciences, C.I.T. Campus, Taramani, 
 Chennai - 600 113, India} 

\begin{abstract}
We study the evolution of a random weighted network with complex 
nonlinear dynamics
at each node, whose activity may cease as a result of interactions with other
nodes. Starting from a knowledge of the micro-level behaviour at each node, we 
develop a macroscopic description of the system in terms of the 
statistical features of the subnetwork of active nodes.
We find the asymptotic characteristics of this subnetwork 
to be remarkably robust:
the size of the active set
is independent of the total number of nodes in the network, and
the average degree of the active nodes is independent
of both the network size and its connectivity. These results suggest
that very different networks evolve to
active subnetworks with the same characteristic features.
This has strong
implications for dynamical networks observed in the natural world, notably
the existence of a characteristic range of links per species across ecological
systems.
\end{abstract}
\pacs{05.45.-a, 89.75.Hc, 89.75.-k}

\maketitle

With the recent surge of interest in complex networks \cite{Newman}, the
behaviour of dynamical units interacting on networks has become a problem
of crucial relevance to areas ranging from physics to biology to engineering.
Coupled nonlinear systems, such as oscillators and maps,
have been extensively investigated on regular lattices \cite{Kan}.
The graph theoretic aspects
of random networks have also received considerable attention \cite{Bollobas}.
However, there have been very few studies on networks with 
nonlinear dynamics at the nodes.
Here we focus on this relatively unexplored area of random networks of 
nonlinear maps, and study
the role played by the network properties on the time-evolution of the
dynamical states of the nodes, in particular, and the global characteristics
of the evolved network, in general.

We consider networks with 
a wide range of ($i$) size (i.e., number of nodes, $N$), 
($ii$) connectivity $C$ between nodes
(i.e., the density of links), ($iii$) measure of interaction
strength $\sigma$ that determines the weights of the connections (i.e., how
strongly the nodes are coupled) and ($iv$) local dynamics at the nodes
(ranging from regular to chaotic).
The important feature here is that even though the isolated nodes may
exhibit a wide range of
activity, the network yields
generically chaotic global dynamics. This can result in 
a fraction of the nodes being driven
to a state of null activity, implying that 
under interactions a certain set of nodes show a transition
from persistent to transient activity. In this paper, we examine the properties
of the {\em subnetwork of nodes with persistent activity}. We show that
the size of this subnetwork of active nodes
is remarkably independent of the network size.
Further, the total number of links
in this subnetwork is independent of both the size and the connectivity
of the network. These results have considerable significance for the 
observable properties of networks occurring in nature.

The work reported in this paper can be seen in the context of deriving
a statistical mechanics of interacting dynamical 
elements \cite{Ker57,Lei65}. Starting from
a micro-level dynamical description, where the relevant variables are the
local states of each node in the network, we would like to achieve a
macroscopic description of the system in terms of the number of active nodes, 
and would like to understand how such macro-variables are determined by 
overall network properties, such as, $N$, $C$ and $\sigma$. 
One would have naively supposed that the macroscopic system variable of 
interest, namely, the size of the persistently active subnetwork, would
be an extensive quantity.
However, our results show that this macroscopic quantity does not
scale with system size. This implies the existence of 
``universal''
relations between various gross network properties in the asymptotic state,
and the emergence of characteristic robust features independent of network size.

Our model is quite general: it has $N$ dynamical elements in a network with 
random nonlocal
connectivity.
The dynamical state of each node $i ( = 1 \ldots N)$ at time $n$
is associated with
a continuous variable $x_i (n)$, which is the microscopic variable of
interest in the system.
The interaction
between two nodes is given by a
coupling coefficient $J_{ij}$. We consider the most general case
where these coefficients can be asymmetric ($J_{ij} \neq J_{ji}$)
and can be either positive or negative.
The time-evolution of the system is given by
\begin{equation}
\label{e.1}
x_i (n+1) = f [ x_i (n) \{ 1 +  \Sigma_j J_{ij} x_j (n) \} ],
\end{equation}
where $f$ represents the local on-site dynamics. In this
paper we have shown representative results for $f$ chosen 
to be the exponential map,
\begin{equation}
\label{e.2}
f (x) = x e^{r(1-x)}, ~{\rm if}~ x > 0;  ~= 0, ~{\rm otherwise},
\end{equation}
$r$ being the nonlinearity parameter leading from periodic behaviour to
chaos \cite{Ric54}.
This belongs to the class of maps defined over the 
semi-infinite interval $[0, \infty]$ rather than a finite,
bounded interval (e.g., as is the case for logistic map).
This allows us to explore arbitrary distributions of couplings between nodes,
unlike maps bounded in an interval, which are well-behaved only for
restrictive coupling schemes. In addition, in our case, 
the nonlinearity parameter $r$ is not artificially restricted by
the domain of definition of the map. All these features increase the
generality of our results, and additionally, such maps provide
a more accurate description of natural processes, e.g., population
dynamics \cite{Has76,Bel81}. 

The connectivity matrix ${\bf J} = \{ J_{ij} \}$ is, in general, 
a sparse matrix,
with probability $1-C$ that an element is zero.
The diagonal entries
$J_{ii} = 0$ indicate that in the absence of interactions,
the local nonlinear map (2) completely determines the dynamical
state of each node. The non-zero entries in the matrix are
chosen from a normal distribution with mean 0 and variance $\sigma^2$.
Note that we have also used uniform distributions over the interval
$[- \sigma, \sigma]$ without any qualitative changes in the results.

Initially, the states of all the $N$ nodes are randomly
distributed about $x = 1$.  
During the evolution of the network, if the state of a node becomes $x \le 0$, 
it stops being active and
subsequently has no interaction with the rest of the network.  
Note that, in the absence of coupling, the maps describing the dynamics 
at individual nodes do not cease activity.
This allows us to focus on the instability induced by network interactions,
rather than the 
intrinsic behaviour of the nodes.
As a result of these interactions, 
the number of persistent nodes (i.e., with $x > 0$) decreases
rapidly from the initial value, but eventually attains a 
steady state.
This is because, at the initial
stages, the population of each node undergoes strong fluctuations
due to interaction with other nodes coupled to it,
resulting in the cessation of activity of a large number of nodes.
Within a very short time,
the effective number of interacting nodes decrease
and, consequently, the intensity of such
fluctuations is also reduced.  We have continued the simulations for
up to $10^4$ iterations, when the probability of further extinctions
was found to become extremely small. We then look at the number of 
nodes which survive with persistent activity as a function of the 
model parameters (Fig. 1).
\begin{figure}
\includegraphics[width=0.48\linewidth,clip]{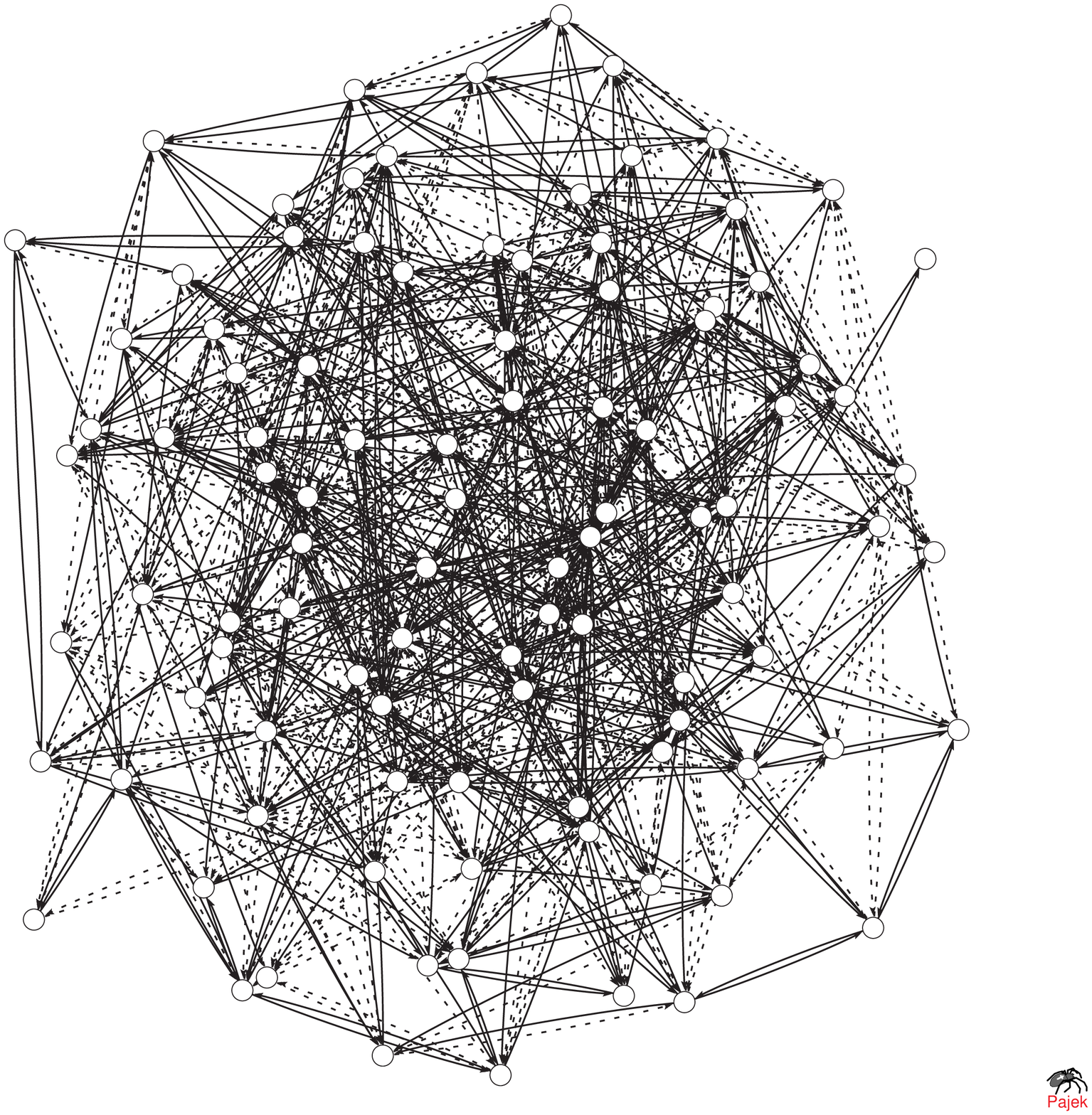}
\includegraphics[width=0.48\linewidth,clip]{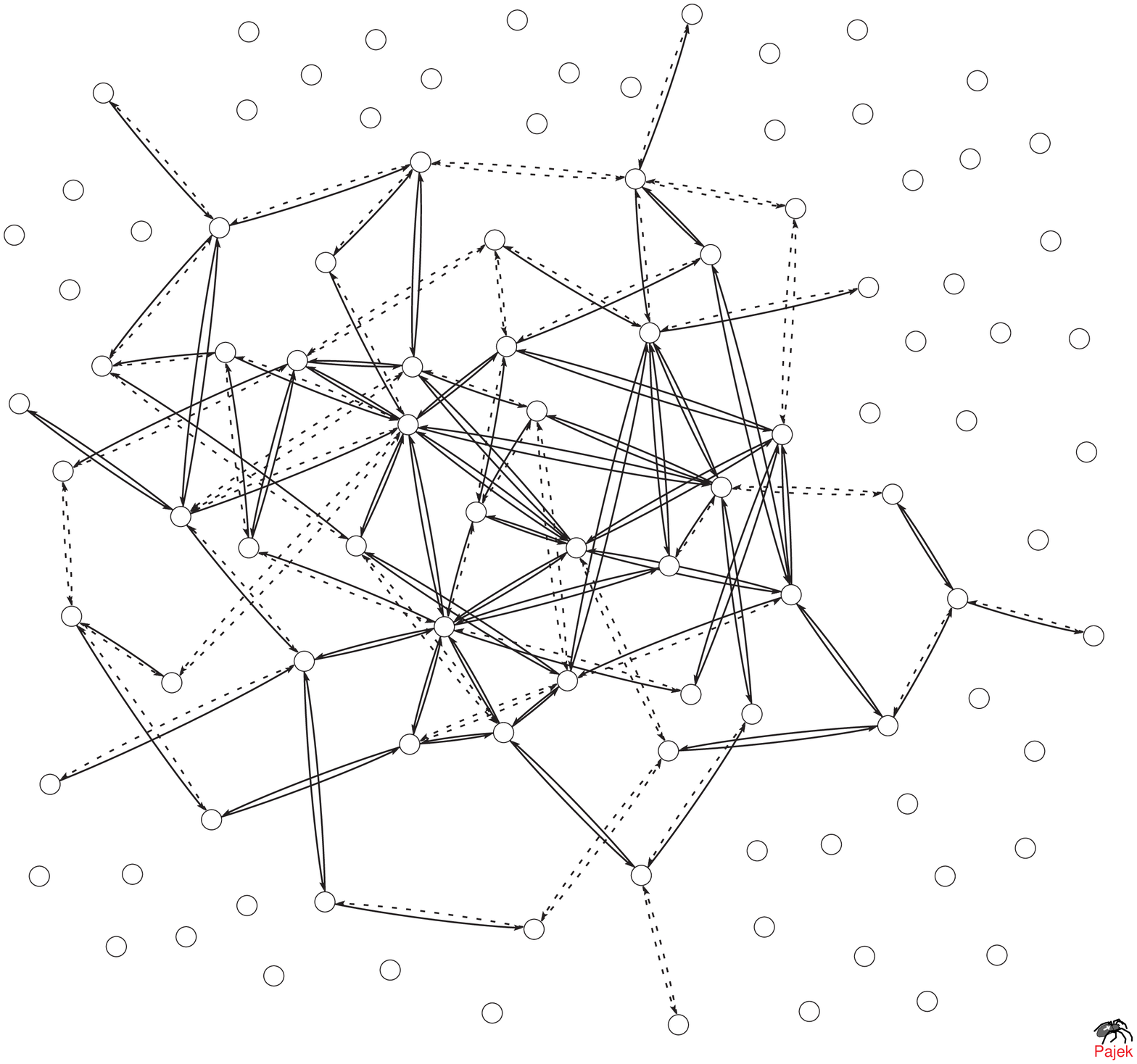}
\caption{The initial (left) and asymptotic (right) state of a
network with $N = 100$, $C = 0.1$,
$\sigma = 0.1$, $r = 4$.
Only the connections between active nodes are shown. Solid
lines represent positive links, while broken lines represent negative links.}
\vspace{-0.5cm}
\label{f.1}
\end{figure}

The number of nodes with persistent activity is a measure of the 
global stability of the network. The information that we get from
this is very different from the local stability and complementary to it.
Note that, in the study of interdisciplinary problems,
global stability (or persistence) is often much more relevant than the
more commonly used measures of
local stability. For example,
networks susceptible to catastrophic failures or crashes are
extremely common in the real world. In such problems, the
quantity of interest is the system's global stability, as reflected in the
survival probability of nodal activity, rather than local stability,
which, in the absence of regular equilibria, does not contribute to our 
understanding of the overall system dynamics \cite{note1}.

We now look at the features of the asymptotic subnetwork consisting of
the nodes which survive with persistent activity. The first significant
feature of this emergent subnetwork is that its size is independent
of the system size $N$. This is clearly evident from Fig. 2, which
shows that the size of the active subnetwork quickly approaches its
asymptotic value $N_{active}$ which is a constant with respect to $N$
(within error bars).
For
example, for the representative case of $\sigma =0.1, C =1, r = 4$, we find
that $N_{active} =7.705 \pm 1.260$ for $N = 250$, 
while for $N=1000$, $N_{active} =7.835 \pm 1.247$.
\begin{figure}
\includegraphics[width=0.85\linewidth,clip]{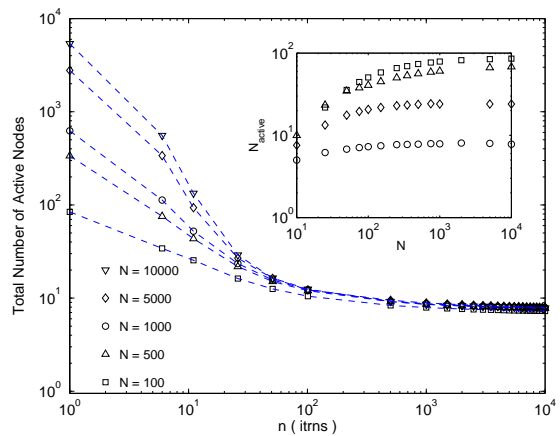}
\caption{
Time-evolution of the number of active nodes for networks
of various sizes $N$ ($C$ = 1, $\sigma$ = 0.1 , $r = 4$), showing the 
rapid decay to the asymptotic value that is independent of $N$.
(Inset) The asymptotic number of 
active nodes, $N_{active}$, as a function 
of network size $N$ ($\square$: $C$ = 0.1, $\sigma$ = 0.1, $r = 4$;
$\bigtriangleup$: $C$ = 1, $\sigma$ = 0.1, $r = 2$;
$\diamond$: $C$ = 0.1, $\sigma$ = 0.5, $r = 4$;
$\circ$: $C$ = 1, $\sigma$ = 0.1, $r = 4$). Note that, in all cases,
$N_{active}$ is independent of the network size $N$, for large $N$.
}
\vspace{-0.5cm}
\label{f.2}
\end{figure}

In the absence of connections ($C=0$), $N_{active}$ ($= N$)
is obviously extensive. But for $C > 0$,  
$N_{active}$ saturates to a value independent of $N$. This non-extensivity
for the active subnetwork has significant implications. For instance, 
let us consider two stable networks, each of which has $N_{active}$
persistently active nodes. On being joined together 
(analogous to two distinct ecological systems being suddenly linked to one
another), the merger initially results in a 
high number of active nodes, which is essentially the sum of the active
nodes of the two components (= $2 N_{active}$). However, 
the new connections prompt a fresh wave of extinctions to occur,
resulting in the number of active nodes 
rapidly settling back to the characteristic value $N_{active}$.

Another remarkable feature is that the average 
number of links per node in the active subnetwork, $k_{active}$,
is independent of $N$, as well as $C$. 
Therefore, 
$k_{active}$ is independent of $k_0 (= N C )$, the average degree of the
entire network (Fig. 3). 
The significance of this result is evident: regardless of the size and
connectivity of the network, the nodes in the active subnetwork has a 
{\em characteristic number of links}.
Together with the earlier result of a characteristic size for
the active subnetwork ($N_{active}$), this implies
that the total number of links ($N_{active} \times k_{active}$) 
connecting the active nodes is a robust quantity.
\begin{figure}
\includegraphics[width=0.85\linewidth,clip]{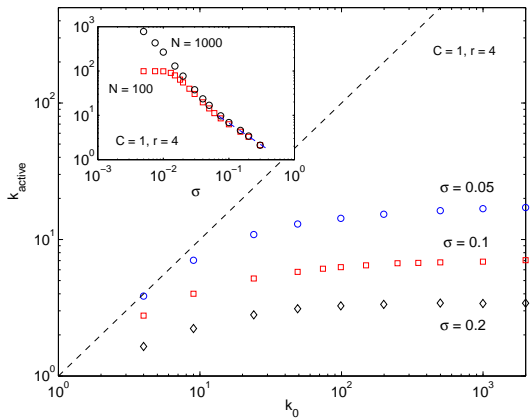}
\caption{
Asymptotic number of links per node in the active subnetwork 
($k_{active}$) as a function of
the number of links per node of the full network ($k_0$) for different 
values of $\sigma$
($C = 1, r = 4$). The broken line represents $k_{active} = k_0$.
The inset shows $k_{active}$ as a function of
$\sigma$ for $N = 100$ and 1000. Note that the two data sets
converge at a high value of $\sigma$ and match well with the broken
line representing $\sim 1/\sigma$.
}
\vspace{-0.5cm}
\label{f.3}
\end{figure}

The active subnetwork is further characterised by disassortativity, i.e.,
nodes with high degree connect preferentially to nodes having fewer links,
a feature observed in many biological and technological networks. For
instance, the disassortativity index as defined by Newman \cite{New02}
is $-0.11$ for $N = 100, C = 0.5, \sigma = 0.1, r = 4$.

The fact that the size of the active set is independent of the
network size $N$ may be naively expected from the May-Wigner
stability criterion \cite{Sin05}. According to this, a system is critically 
stable if $N C \sigma^2 \simeq 1$. So the size of the asymptotically stable
set $N_{active}$ is proportional to $1/C \sigma^2$. However, 
note that, in an evolving system, the values of $C$ and $\sigma$ for the 
active set also change significantly over time, 
with nodes becoming inactive and their links 
becoming non-functional. 
As a result, we have an interplay between the size and structure of the
subnetwork of active nodes: the size of this set
changes due to the
stability criterion involving the structural parameters $C$ and $\sigma$ 
at that point in time,
while the latter ($C$, $\sigma$) themselves change due to the reduction
in the number of active nodes.
For the May-Wigner argument to hold,
it is crucial that $C$ and $\sigma$ 
eventually attain their own asymptotic values
$C_{active}$ and $\sigma_{active}$, and that these 
are independent of network size $N$. 
To verify this we now look at these properties in the asymptotic
set of active nodes. 

\begin{figure}
\includegraphics[width=0.85\linewidth,height=0.67\linewidth,clip]{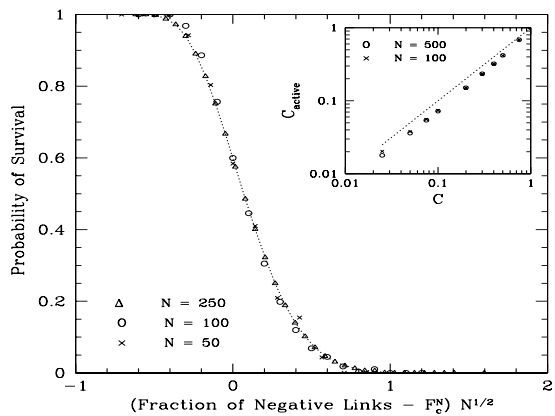}
\caption{
The persistence probability (i.e., survival of activity) of individual
nodes as a function of the fraction of negative links to that node, 
appropriately
scaled by the network size $N$, so that data for different values
of $N$ ($\bigtriangleup$: $N = 250$, $\circ$: $N = 100$, $\times$: $N = 50$) 
collapse
on the same curve using $F_c^{N} = 0.1$ for $N = 50$, $= 0.06$ for $N = 100$
and $= 0.039$ for $N = 250$. The network parameters are $C = 
0.1$, $\sigma = 0.1$ and $r = 4$. The inset shows the connectivity of the
active subnetwork,
$C_{active}$, as a function of $C$ for two 
different network sizes.
}
\vspace{-0.5cm}
\label{f.4}
\end{figure}

First, we examine how the effective connectivities of the asymptotic
subnetwork, $C_{active}$, is different from the network connectivity, $C$. 
Remarkably, $C_{active}$ is found to be independent of the network size
for large $N$. For $0 < C < 1$, 
the effective connectivity of the asymptotic subnetwork
shows a clear trend of evolving to a slightly lower value compared to $C$
[Fig. 4 (inset)], with the
deviation increasing with average interaction strength,
$\sigma$ and local nonlinearity parameter, $r$.
The independence of $C_{active}$ from $N$ can also be 
inferred from Fig. 3, by noting that the $k_{active}$ vs $k_0$ 
curve can be
constructed entirely by considering $C = 1$, for which
$C_{active} (= C)$ is obviously independent of $N$. The same curve
holds for other values of $C$ and therefore one can conclude that $C_{active}$
is independent of $N$, for all $C$. Further, 
$N_{active} \sim 1/C_{active}$ \cite{Sin05}, 
which is consistent with our earlier result
that $N_{active}$ is independent of $N$.

Next, we address the question of the independence of $\sigma_{active}$ with
respect to $N$, by looking at
the distribution of the connection weights $J_{ij}$ of the persistent
nodes. Specifically, we examine how this differs from the distribution 
of $J_{ij}$ in the full network.
We observe that starting from a Gaussian distribution (for instance), the
distribution that emerges is independent of $N$, 
indicating
that $\sigma_{active}$ is independent of the network size. Further, 
the distribution is markedly skewed towards positive weights 
implying a selective tendency of nodes with high negative links
to be eliminated. This is consistent with
the probability of survival of a {\em single} node being
a decreasing function of the relative number of its negative links (Fig. 4).

\begin{figure}
\includegraphics[width=0.85\linewidth,height=0.67\linewidth,clip]{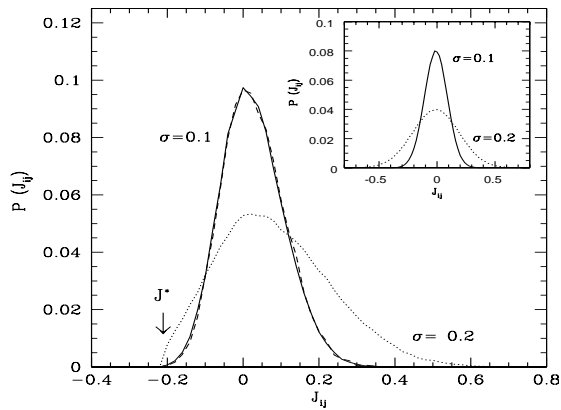}
\caption{Distribution of
the (non-zero) interaction strengths $J_{ij}$
between active nodes of a network for $C = 0.5, \sigma = 0.1$ (solid line),
$C = 1, \sigma = 0.1$ (dashed line) and $C = 0.5, \sigma = 0.2$ (dotted line). 
Note that, the cutoff at the lower end, $J^*$, is independent of $\sigma$.
The inset shows the
$J_{ij}$ distribution in the full network for the same set of parameters.
}
\label{fig5}
\end{figure}

To explain the numerical results that show the average degree
of the active subnetwork, $k_{active}$, evolving to a characteristic value, 
we recall that the
average number of links per node  is given by the product of 
$N_{active}$ and 
$C_{active}$. The $C_{active}$ does
not vary too much from $C$, but $N_{active}$ varies significantly from $N$.
So, the number of links essentially is dependent only on $N_{active}$.
If $N_{active}$ settles down to the same constant value independent of $N$, 
it implies that the 
number of links is also a robust quantity (Fig. 3).

Further, $k_{active}$ varies with the overall network average interaction
strength as $\sim 1/\sigma$ [Fig. 3 (inset)]. This can be understood from the 
condition determining persistence of activity for an individual node $i$:
$P ( \sum_j J_{ij} x_j < -1)$. Note that, the contributing terms in the sum 
are due to active nodes which have outgoing links (with non-zero weights) to
node $i$, i.e., the degree of the node in the active subnetwork.
From Fig.~\ref{fig5}, it is apparent that
the asymptotic distribution of $J_{ij}$ for the set of active nodes has a 
positive mean, $\mu$. This implies that the quantity $\sum_j J_{ij} x_j$
has a standard deviation, which is dominated by the leading term,
$\mu k_{active}$, implying $k_{active} \sim 1/\mu$. 
To see how $\mu$ is related to $\sigma$, we look in detail at the
asymptotic distribution of $J_{ij}$. It is immediately apparent that the
dominant change in the asymptotic distribution, vis-a-vis the original
distribution is the loss of the strong negative links. This results in 
the final distribution being truncated at the negative end, with the
cutoff $J^*$ apparently independent of $\sigma$ (Fig.~\ref{fig5}). 
Assuming that the
positive part of the distribution is almost unchanged from its original form,
we see that $\mu = \int_{J^*}^{\infty} J_{ij} P ( J_{ij} )$ essentially
goes as $\sigma$, as the shift of the mean from zero is entirely determined
by the width of the original distribution.

%
In conclusion, we have shown that a very simple model, with very few 
assumptions regarding the node properties
and their dynamics, yields surprisingly robust macroscopic
features of the emergent active system:
(a) the asymptotic number of active nodes is independent
of the network size, and (b) the asymptotic number of links between the 
active nodes is independent
of both the size of the network and its connectivity.
The link removal process is not 
guided here by any explicit criterion designed to achieve a desired end-state
but emerges naturally from the dynamics at the nodes.

The observed non-extensivity of the active subnetwork indicates that
designing robust structures simply by increasing
the redundancy of nodes, keeping the connectivity and interaction strength
distribution unchanged, is not a good strategy, as the number of 
asymptotically active nodes
is independent of the initial number of nodes that one
starts out with. This provides an explanation for similar observations in
natural and artificial 
complex systems, such as the conservation of the number of species in an 
ecosystem
after major extinctions (e.g., after the eruption in Krakatoa) or 
migrations (e.g., after the linking of North and South America) \cite{May78}, 
as 
well as the existence of a characteristic range of links per species (3 to 5)
across different environments \cite{May88}. 

\vspace{-0.5cm}
%


\begin{thebibliography}{99}

\bibitem{Newman} {M.~E.~J.~Newman, SIAM Rev. {\bf 45}, 167 (2003).}
\bibitem{Kan} {K.~Kaneko (Ed.) {\it Theory and Applications of Coupled Map 
Lattices}, (John Wiley, New York, 1993).}
\bibitem{Bollobas} {B.~Bollobas {\it Random Graphs}, (Academic Press, London,
1985).}
\bibitem{Ker57} {E.~H.~Kerner, Bull. Math. Biophys. {\bf 19}, 121 (1957).}
\bibitem{Lei65} {E.~G.~Leigh, Proc. Natl. Acad. Sci. (USA) {\bf 53}, 777
(1965).}
\bibitem{Ric54} {W.~E.~Ricker, J. Fish. Res. Board Can. {\bf 11}, 559 (1954).}
\bibitem{Has76} {M.~P.~Hassell, J.~H.~Lawton and R.~M.~May, J. Anim. Ecol. 
{\bf 45}, 471 (1976).}
\bibitem{Bel81} {T.~S.~Bellows, J. Anim. Ecol. {\bf 50}, 139 (1981).}
\bibitem{note1} {We would like to point out that the dynamics of the 
asymptotic network
does not lead to a fully, or even, partially synchronised state between
the nodes. Indeed there is no global equilibrium, which is verified
by calculating the spectrum of lyapunov exponents of the system.
The spectrum shows strong
spatiotemporal chaos in the global dynamics with no apparent
synchronisation between the nodes with persistent activity.
Even when the local dynamics is regular, the global dynamics of the coupled
system does show chaos.} 
\bibitem{New02} {M.~E.~J.~Newman, Phys. Rev. Lett. {\bf 89}, 208701 (2002).}
\bibitem{Sin05} {S.~Sinha and S.~Sinha, Phys. Rev. E {\bf 71}, 020902 (2005).}
\bibitem{May78} {R.~M.~May, Sci. Am. {\bf 239}(3), 161 (1978).}
\bibitem{May88} {R.~M.~May, Science {\bf 241}, 1441 (1988).}
\end{thebibliography}
\end{document}